# Geographical Analysis: from Distance-based Space to Dimension-based Space

Yanguang Chen

(Department of Geography, College of Urban and Environmental Sciences, Peking University, Beijing 100871, P.R.China. E-mail: chenyg@pku.edu.cn)

**Abstract**: The traditional concept of space in geography is based on the notion of distance. Where there is a spatial analysis, there is a distance measurement. However, the precondition for effective distance-based space is that the geographical systems have characteristic scales. For a scale-free geographical system, the spatial structure cannot be validly described with pure distance, and thus the distance-based space is ineffective for geographical modelling. In the real geographical world, scale-free patterns and processes are everywhere. We need new notion of geographical space. Using the ideas from fractals and scaling relations, I propose a dimension-based concept of space for scale-free geographical analysis. If a geographical phenomenon bears characteristic scales, we can model it using distance measurement; if a geographical phenomenon has no characteristic scale, we will describe it using fractal dimension, which is based on the scaling relations between distance variable and the corresponding measurements. In short, geographical space fall into two types: scaleful space and scale-free space. This study shows a new way of spatial modeling and quantitative analyses for the geographical systems without characteristic scale.
**Key words**: Geographical modeling; Spatial analysis; Characteristic scale; Scaling; Fractals; Geographical space

# 1 Introduction

Geography is a science on the spatial distribution of human and physical phenomena in the world. Geography doesn't care about matter and energy. It is concerned with the non-uniform distribution of matter and energy on the earth's surface in time and space. This is to say, geographical space is heterogeneous space. Spatial heterogeneity indicates regional difference, which in turn indicates spatial information. Geography is a science of spatial information (Chen, 1994; Goodchild, 1992). To research geographical problems, we have to gain, process, and analyze spatial data, describe geographical spatial phenomena, and reveal the hidden order in space and place. The basic and important measurement for spatial description is distance. In this sense, geography is regarded as a discipline in distance (Johnston, 2003; Watson, 1955). Based on distance, various mathematical models and spatial statistic methods for spatial analysis have been developed. With the mathematical models and statistical technique, we can carry out the explanation and prediction for geographical evolution. Unfortunately, in many cases, the explanation and prediction based on these geographic mathematical models and statistics are not accurate and reliable (Portugali, 2000). Maybe there is something wrong with these models and statistics, but what is the problem?

In fact, conventional mathematical modeling and quantitative analysis are based on characteristic scales. The precondition of effective spatial analyses is to find typical numbers to represent characteristic length of a geographical phenomenon. The number may be a determinate length, area, volume, density, eigenvalue, average value, or standard deviation. If we can find out characteristic scales for a geographical system, we can describe and further understand it. However, in many cases, it is impossible for us to find an invalid characteristic scale for a geographical system (Chen, 2008). The reason may be due to the limitations of technology and methods. As a matter of fact, many geographical phenomena have no characteristic scale at all. If a geographical phenomenon bear no characteristic scale, the spatial analytical methods based on distance is ineffective. In this instance, the spatial analyses based on characteristic scales should be replaced by scaling, and the space concept based on distance should be replaced by the space concept based on fractal dimension. Concretely speaking, the geographical analysis based on Euclidean space should be replaced by fractal space. This paper is devoted to distinguishing dimension-based space from distance-based space in geography. In Section 2, the geographical space concept based distance is illuminated by means of distance-decay law; In Section 3, the geographical space concept based fractal dimension is illustrated through spatial scaling law. In Section 4, the integrated spatial analyses by combining distance-based space and dimension-based space are discussed. Finally, the discussion is concluded by summarizing the main points of this work.

# 2 Distance-based geographical space

## 2.1 Distance-decay law

Geographical systems are different from classical physical systems. There is no iron law in geography. However, there are some mathematical models that appear frequently in geographical analysis. These models reflects several basic laws in geographical systems. The most important three ones include *distance-decay law*, *rank-size law*, and *allometric growth law*. In the set of these mathematical laws of geography, the most significant one is the distance-decay law. The important spatial analytical models and methods, including gravity models, spatial interaction modeling, and spatial autocorrelation analysis, are all based on the distance-decay principle. The so-called first law of geography presented by Tobler (1970; 2004), which reads "everything is related to everything else but near things are more related than distant things", are actually based on distance-decay law. The spatial allometric growth law can be associated with distance-decay law directly, and the rank-size law can be linked to the distance-decay law indirectly. A number of functions can be employed to characterize distance decay effect in geographical world (Table 1). Among various distance decay functions, two ones are typical and in common use. One is negative exponential function, and the other, inverse power function.

Table 1 General forms of distance decay functions in geography

| Type | Name | Function | Parameter |
|---|---|---|---|
| Single logarithm model | Normal (Gauss function) | $f(r) = f_0 e^{-br^2}$ | $f_0, b$ |
| | Exponential (I) | $f(r) = f_0 e^{-br}$ | $f_0, b$ |
| | Square root exponential | $f(r) = f_0 e^{-br^{1/2}}$ | $f_0, b$ |
| | Logarithmic | $f(r) = f_1 - b\ln(r)$ | $f_1, b$ |
| Hybrid model | Exponential (II) | $f(r) = f_1 e^{b/r}$ | $f_1, b$ |
| | Lognormal | $f(r) = f_1 e^{-b(\ln(r))^2}$ | $f_1, b$ |
| | Gamma | $f(r) = f_1 r^{-a} e^{-br}$ | $f_1, a, b$ |

| | | | | |
|---|---|---|---|---|
| Double logarithm model (power law) | Pareto | $f(r) = f_1 r^{-a}$ | $f_1, a$ | |

**Source:** Chen (2010); Haggett, *et al* (1977, page31); Taylor PJ (1975); Zhou (1995, page 360). **Symbols**: $f(r)$=action or interaction strength; $r$=distance; $a$, $b$, $f_0$, $f_1$=constants, $e$=exponential constant (2.7183); ln=natural logarithm function.

## 2.2 Geographical based on distance

In a sense, the conventional space of geography is a type of distance-based space. In geography, spatial analysis is mainly based on distance variable (Johnston, 2003). Watson (1955) once pointed out: "Distance, as a measurable phenomenon, is basic to the study of geography. When a geographer observes a fact and locates it as part of the earth's scene, he expresses that location as distance from the prime meridian and the equator. " The important models and analytical methods are directly or indirectly associated with distance variable and distance-decay functions (Table 2). Typical models include the gravity models (Carey, 1858; Chen, 2015a; Converse, 1949; Fotheringham and O'Kelly, 1989; Ravenstein, 1885; Reilly, 1931; Rybski *et al*, 2013; Sen and Smith, 1995), urban density models (Clark, 1951), traffic network density models (Smeed, 1963), spatial interaction models (Wilson, 1968; Wilson, 1970; Wilson, 2000; Wilson, 2010), spatial autocorrelation analyses (Anselin, 1995; Cliff and Ord, 1973; Cliff and Ord, 1981; Geary, 1954; Getis, 2009; Getis and Ord, 1992; Moran, 1950), and spatial auto-regression analysis or spatial lag regression models (Anselin, 1988; Ward and Gleditsch, 2008), and son. The well known central place theory are spatial hierarchy based on distance (Batty and Longley, 1994; Christaller, 1933; Lösch, 1940).

Table 2 Commonly used geographic mathematical models for spatial analysis and their spatial properties

| Model | Distance decay | Function | Characteristic parameter | Space nature |
|---|---|---|---|---|
| Density distribution | Exponential | $\rho(r) = \rho_0 e^{-r/r_0}$ | $r_0$ | Based on characteristic distance |
| | Power law | $\rho(r) = \rho_1 r^{-a}$ | $a$ | Based on scaling exponent |
| Gravity | Exponential | $F_{ij}(r) = GQ_iQ_j e^{-r_{ij}/r_0}$ | $r_0$ | Based on characteristic distance |
| | Power law | $F_{ij}(r) = GQ_iQ_j r_{ij}^{-a}$ | $a$ | Based on scaling exponent |
| Spatial interaction | Exponential | $T_{ij} = A_i B_j O_i D_j e^{-r_{ij}/r_0}$ $\begin{cases} A_i = 1/\sum_j B_j D_j e^{-r_{ij}/r_0} \\ B_j = 1/\sum_i A_i O_i e^{-r_{ji}/r_0} \end{cases}$ | $r_0$ | Based on characteristic distance |
| | Power law | $T_{ij} = A_i B_j O_i D_j r_{ij}^{-a}$ $\begin{cases} A_i = 1/\sum_{j=1}^n (B_j D_j r_{ij}^{-a}) \\ B_j = 1/\sum_{i=1}^n (A_i O_i r_{ji}^{-a}) \end{cases}$ | $a$ | Based on scaling exponent |
| Spatial autocorrelation | Exponential | $I = z^T W z$ $\begin{cases} W = [w_{ij}] = \left[v_{ij}/\sum_{i=1}^n \sum_{j=1}^n v_{ij}\right] \\ v_{ij} = e^{-r_{ij}/r_0} \end{cases}$ | $r_0$ | Based on characteristic distance |
| | Power law | $I = z^T W z$ $\begin{cases} W = [w_{ij}] = \left[v_{ij}/\sum_{i=1}^n \sum_{j=1}^n v_{ij}\right] \\ v_{ij} = r_{ij}^{-a} \end{cases}$ | $a$ | Based on scaling exponent |
| Spatial auto-regression | Exponential | $y = \mu + \beta W y + \varphi x + \gamma W x$ $\begin{cases} W = [w_{ij}] = \left[v_{ij}/\sum_{i=1}^n \sum_{j=1}^n v_{ij}\right] \\ v_{ij} = e^{-r_{ij}/r_0} \end{cases}$ | $r_0$ | Based on characteristic distance |
| | Power law | $y = \mu + \beta W y + \varphi x + \gamma W x$ $\begin{cases} W = [w_{ij}] = \left[v_{ij}/\sum_{i=1}^n \sum_{j=1}^n v_{ij}\right] \\ v_{ij} = r_{ij}^{-a} \end{cases}$ | $a$ | Based on scaling exponent |

**Note on symbols**: $\rho(r)$=density; $r$=distance; $a$, $b$, $G$, $\rho_0$, $\rho_1$, $\mu$, $\beta$, $\varphi$, $\gamma$=constants; $e$=exponential constant (2.7183); $F_{ij}$=gravity force; $Q$=size; $T_{ij}$=flow quantity from region $i$ to region $j$; $O_i$=inflow quantity; $D_j$=inflow quantity; $A_i$, $B_j$=scaling factor for spatial interaction; $I$=Moran's index; $v_{ij}$= spatial contiguity; $w_{ij}$= spatial weight; $x$=independent variable; $y$=dependent variable; $z$=standardized $x$ or $y$; $n$=region number, $i,j$=1,2,…,$n$.

## 2.3 Difficulty of spatial measurements

Geographic systems are complex spatial systems. We can study this kind of systems with the help of maps. To map geographical phenomena is essentially to construct a model (Holland, 1998). In fact, various elements in a geographic system can be abstracted into points, lines and areas on a map (Table 3). To describe a point, we should know its location, but if we want to describe two or more points, we should know the distance between any two points besides locations. To describe a line, we should know its length, which is equivalent to a distance. To describe an area, we should know its size, and the radius of the equivalent circle of the area is also a distance. In a word, in geographical analysis, distance always appears directly or indirectly everywhere.

**Table 3 Points, lines, and area in geographical models**

|  | Point | Line | Area |
|---|---|---|---|
| Point | e.g., cities in a region | e.g., cities along a river | e.g., a city and its hinterland |
| Line |  | e.g., network of roads and railways | e.g., traffic network within an urbanized area |
| Area |  |  | e.g., urban domain of attraction area trade area |

Unfortunately, when geographers try to accurately measure the length of a geographical line or the size of a geographical area, they often fall into dilemma of spatial measurement. The typical problem is what is called "conundrum of length" in geography (Batty, 1991). Geographers found that measured length of a geographical line such as a river increases with increasing accuracy of measurement (Goodchild and Mark, 1987; Nystuen, 1966). This is termed *Steinhaus paradox* (Batty, 1991; Bibby, 1972; Coffey, 1981; Goodchild and Mark, 1987). The phenomenon of scale dependence of the length of an irregular curves such as a river channel was earlier discussed by Steinhaus (1954, 1960). Back then, geographers were only one step away from discovering fractal phenomena. Haggett and Chorley (1969, page 67) observed: "the more accurate an empirical line is measured, the longer it gets". But the geographers' judgments were denied by the statistics at that time. Statistician Bibby (1972) make a comment as below: "Thus stated, the observation is correct. That 'at the molecular level the length approaches infinity' is false, since a variable can increase and yet have a finite upper bound." The inference of Bibby (1972) was based on a geometric series, which is equivalent to the cumulative result of exponential decay function. During this period, Mandelbrot (1969) used power laws to analyze geographical lines such as coastlines and found a scientific solution to the conundrum of length. In fact, not only coastlines and rivers, but also other geographical lines such as border lines, urban boundary lines, traffic lines, ridge lines, and so on, all involve scale dependence (Batty and Longley, 1994; Longley and Batty, 1989a; Longley and Batty, 1989b; Mandelbrot, 1982). In short, geographical lines are always fractal lines or fractal-like curves with scaling symmetry. Where China's Yangtze River, the great wall and the provincial boundary line are concerned, the results of the newer measurements are always greater than the older results in previous measurements. The reason lies in the increasing accuracy of measurement. The more accurate the measurement, the smaller the space yardstick used for the measurement. Not only geographical lines, but also the area of a region is sometimes impossible to be certainly measured. For example, the size of China's territory depends on the measurement scale, and this involves more complex geographical fractal analysis (Chen, 2012).

One of the ways of spatial sampling and measurement is geographical division. In many cases, we have to make use of zonal systems to measure geographical phenomena and obtain spatial data. A zonal system has a great many spatial units, which bear different areal sizes (Batty and Longley, 1994). A significant problem is that the structure of a zonal system influences the results of spatial measurement and statistical analyses. This involves so-called modifiable areal unit problem (MAUP), which was found by Gehlke and Biehl (1934) and rediscovered by Openshaw (1983) and Arbia (1988). MAUP implies that the size and shape of area units in a zonal system affects the calculated results and statistical inference conclusion. These years, MAUP has become a hot topic and well-known difficult problem in geography. However, geographers have different views on the essence of MAUP (Cressie, 1996; Kwan, 2012; Swift *et al*, 2008; Unwin, 1996; Viegas *et al*, 2009). To solve MAUP, we should make use of the idea from fractals and scaling. Due to scale dependence, the spatial measurements are often uncertain. If we utilize box-counting method to replace arbitrary zonal systems, we will be able to avoid a number of MAUP. The box-counting method are based on scaling process rather than characteristic scales.

# 3 Dimension-based geographical space

## 2.1 Spatial scaling law

Scaling is essentially invariance under contraction or dilation transformation. If a geographical model bears invariance under spatial contraction or dilation, it is regarded as following spatial scaling law. The invariance under a transform represents a type of symmetry (Mandelbrot, 1982). If a spatial phenomenon bears scaling nature, the distance-based space will be invalid for geographical analysis and should be replaced with dimension-based space. The dimension-based space denotes the geographical space based on fractal dimension or generalized fractal dimension. In mathematics and science, dimension is utilized in describing spatial concepts such as points, lines, areas, and volumes. In empirical studies, a dimension implies a measurement such as length, width, or height. If a geographer talks about the dimensions of a space or place, he is referring to its size, shape, and proportions. In short, in scientific research, dimension is a spatial characteristic. The dimension of Euclidean geometry is known and has no information: point is 0 dimension, line is 1 dimension, face is 2 dimension, body is 3 dimension. Information lies in uncertainty. If a quantity is known without measurement, it gives no information. Therefore, the Euclidean dimension generally does not have much information. Fractal dimension needs to be measured in order to know the specific value, including spatial information. After the emergence of fractal geometry, dimension entered the empirical science from the theoretical science. Fractal dimension is the basic and important parameter for scaling analysis in geography.

Scaling analysis cannot be applied to scaleful phenomena, just as conventional mathematical tools is generally not suitable for scale-free phenomena. Using contraction or dilation transform, we can test whether or not a model bear scaling property. For a function $f(x)$, suppose that it satisfies the following relation

$$\mathbf{T}f(x) = f(\xi x) = \xi^a f(x) = \lambda f(x), \qquad (1)$$

where $x$ refers to an argument, $\mathbf{T}$ represents a scaling transform, i.e., contraction or dilation transform, $\xi$ is a scale factor for the scaling transform, $a$ is a scaling exponent, and $\lambda = \xi^a$ is the eigenvalue of the transform $\mathbf{T}$. The eigenvalue is a function of the scaling exponent, and the scaling exponent is always associated with fractal dimension. For example, applying scaling transform to the gravity model based on power-law decay yields

$$\mathbf{T}F_{ij}(r) = F_{ij}(\xi r) = GQ_iQ_j(\xi r_{ij})^{-a} = \xi^{-a}GQ_iQ_jr_{ij}^{-a} = \lambda F_{ij}(r), \quad (2)$$

which satisfies the scaling relation, equation (1). The distance decay exponent, $a$, proved to be associated with fractal dimension (Chen, 2015a). This indicates that the gravity model based on inverse power law follows scaling law. In contrast, applying the contraction-dilation transform to the gravity model based on exponential decay function yields

$$\mathbf{T}F_{ij}(r) = F_{ij}(\xi r) = GQ_iQ_je^{-(\xi r_{ij})/r_0} = (GQ_iQ_j)^{1-\xi}F_{ij}(r)^\xi \neq \lambda F_{ij}(r), \quad (3)$$

which does not satisfy the scale invariance relation, equation (1). This implies that the gravity model based on negative exponential decay disobeys the scaling law. However, applying a translational transform to

$$\mathbf{T}^*F_{ij}(r) = F_{ij}(r+\varsigma) = GQ_iQ_je^{-(r_{ij}+\varsigma)/r_0} = e^{-\varsigma/r_0}(GQ_iQ_je^{-r_{ij}/r_0}) = \lambda F_{ij}(r), \quad (4)$$

where $\mathbf{T}^*$ denotes translational transform, $\varsigma$ refers to translation scale, and $\lambda=\exp(-\varsigma/r_0)$ is the eigenvalue of the translation transform $\mathbf{T}^*$. Equation (2) suggests spatial scaling symmetry, indicating invariance understand spatial contraction and dilation. In contrast, equation (4) suggests spatial translation symmetry, indicating invariance understand spatial translation.

## 2.2 Geographical spatial based on dimension

Geographical phenomena seem to be randomly distributed, but they contain spatial order in deep structure. Geographical systems follow scaling law in many aspects. As indicated above, there are three significant mathematical laws in geography, that is, distance-decay law, rank-size law, and allometric growth law. Each law involves a number of mathematical models (Table 4). These models fall into two categories: one is those based on scaleful decay functions such as negative exponential function, the other is based on scale-free decay function, i.e., inverse power law (Table 1). If the distance decay functions do not follow scaling law, which is formulated as equation (1), the distance effect bears characteristic scales, and belongs to scaleful decay. This type of spatial processes belong to distance-based on space and can be described, modeled, and analyzed using conventional mathematical methods. In contrast, if the distance decay functions follow scaling law, the distance effect possesses no characteristic scale, and belongs to scale-free decay. This type of spatial processes belong to dimension-based space and cannot be characterized, modeled and examined using traditional mathematical methods. We need new mathematical tools such as fractal geometry, allometric theory, complex network theory, wavelet analysis, renormalization group, and so on. In the distance-based geographical space, the quantitative analyses are based on characteristic length associated with distance. However, in the dimension-based geographical space, the quantitative analyses should be based on scaling exponents, which is directly or indirectly associated with fractal dimensions.

**Table 4 Three basic laws in human geography: distance decay, rank-size distribution, and allometric growth**

| Law | Model | Space and data |
| --- | --- | --- |
| **Distance decay law** | Density distribution models | Real space: The basic model is based on spatial series data |
|  | Gravity models |  |
|  | Spatial interaction models |  |
|  | Spatial autocorrelation models |  |
| **Rank-size law** | Zipf's law | Order space: The basic model is based on hierarchical series data |
|  | Pareto distribution |  |
|  | Davis' $2^n$ rule |  |
| **Allometric growth law** | Urban area and population size allometry model | Phase space: The basic model is based on temporal series data. |
|  | Urban area and perimeter length allometry model |  |
|  | Central city and urban system allometry model |  |

Mathematical methods are often based on invariance under a transform and commensurability in the invariance. Applying a transform, **T**, to a function, which acts as a geographical model. If the result of transformation is linearly proportional to the original function, we will say that the function does not change in structure after going through the transform. The function can be regarded as eigenfunction of the transform, and the proportionality coefficient is the corresponding eigenvalue (Prigogine and Stengers, 1984). Many quantitative analyses rely heavily on the eigenvalues (Chen, 2008). For example, according to equation (4), the gravity model based on exponential decay satisfies the invariance under translational transform, and eigenvalue is associated with the characteristic length $r_0$. In contrast, according to equation (2), the gravity model based on power-law decay takes on the invariance under scaling transform, and eigenvalue is associated with the scaling exponent, *a*. Power laws are a kind of indication of scaling in geographical systems. The appearance of a power law usually implies the existence of scaling.

Scaling analyses depend mainly on scaling exponents, which proved to be associated directly or indirectly with fractal dimension. Let's see a number of typical geographical models. If urban density distribution follows the inverse powers law, then the cumulative distributions will follow power laws and we have

$$A(r) = A_0 r^{D_a}, \tag{5}$$

$$P(r) = P_0 r^{D_p}, \tag{6}$$

where $r$ denotes the distance from city center, $A(r)$ is the land use area within the circle with a radius $r$, $P(r)$ is the population quantity within the circle with a radius $r$, $A_0$ and $P_0$ are proportionality coefficients, $D_a$ is the fractal dimension of urban land use form, and $D_p$ is the fractal dimension of urban population distribution. The two fractal dimension values are based on distance $r$ and be expressed as $D_a^{(r)}$ and $D_p^{(r)}$. Combining equations (5) and (6) yields

$$\frac{A(r)}{A_0} = \left(\frac{P(r)}{P_0}\right)^{D_a/D_p}, \tag{7}$$

which reflects the spatial allometric scaling relation between urban area and urban population size (Chen *et al*, 2019). Equations (7) can be rewritten as below

$$A(r) = aP(r)^b = aP(r)^{D_a^{(r)}/D_p^{(r)}}, \tag{8}$$

in which $a$ is proportional constant, and $b$ is the scaling exponent. The parameters can be expressed as follows

$$a = A_0 P_0^{-D_a^{(r)}/D_p^{(r)}}, \tag{9}$$

$$b = \frac{D_a^{(r)}}{D_p^{(r)}}. \tag{10}$$

The second parameter is related to an eigenvalue. Conclusions can be reaches as follows. First, allometric scaling law can be derived from distance decay laws. Second, the allometric scaling exponent is just the ratio of the fractal dimension of urban land use form to the fractal dimension of urban population distribution.

The spatial allometric scaling can be converted into temporal allometric scaling and hierarchical allometric scaling. Replacing the distance variable, $r$, in equation (8) with time variable, $t$, yields a longitudinal allometric relation as below

$$A(t) = aP(t)^b = aP(t)^{D_a^{(t)}/D_p^{(t)}}, \quad (11)$$

which indicates the dynamic allometric process in time direction. This is the allometrtic growth model in a narrow sense. Substituting the distance variable in equation (8) with rank variable, $k$, yields a transversal allometric relation as follows

$$A(k) = aP(k)^b = aP(k)^{D_a^{(k)}/D_p^{(k)}}, \quad (12)$$

which indicates the cross-sectional allometric scaling in rank-size direction. This is the allometrtic growth model in a broad sense. The cross-sectional allometric model can be associated with the rank-size law. The well-known Zipf's law is often expressed as

$$P(k) = P_1 k^{-q}, \quad (13)$$

where $k$ refers to rank, $P(k)$ denotes the corresponding city population size, $P_1$ is the proportionality coefficient indicating the largest size, and $q$ is the Zipf exponent, namely, the scaling exponent of the rank-size distribution. Substituting equation (13) into equation (12) yields

$$A(k) = a(P_1 k^{-q})^b = aP_1^b k^{-bq} = A_1 k^{-p}, \quad (14)$$

in which $A(k)$ denotes the urban area of the $k$th city, $A_1 = aP_1^b$ is the proportionality coefficient indicative of the largest size, and $p = bq$ is another Zipf exponent. This suggests that if urban population size distribution follow Zipf's law, the corresponding urban areal size distribution also follow Zipf's law (Chen, 2008). In terms of equation (10), the scaling exponent can be expressed as follows

$$p = bq = q \frac{D_a^{(k)}}{D_p^{(k)}}. \quad (15)$$

Thus we have a proportional relation as below:

$$\frac{p}{q} = \frac{D_a^{(k)}}{D_p^{(k)}}, \quad (16)$$

This implies that the ratio of two allometric scaling exponents is equal to the ratio of two fractal dimensions (Chen, 2014).

# 4 Integrated spatial analysis

## 4.1 Two types of geographical phenomena

Geographical phenomena can be divided into two types: one is the phenomenon with characteristic scales, the other is the phenomenon without characteristic scale. The former can be termed scaleful geographical phenomena, and the latter is termed scale-free phenomena (Chen, 2015a; Chen, 2015b). If a geographical phenomenon bears characteristic scales, it has determinate length, area, volume, eigenvalue, average value, or standard deviation. The probability density distribution of a scaleful phenomenon often takes on a unimodal curve (e.g., gamma curve), or the curve can be converted into a unimodal curve (e.g., exponential decay curve). In contrast, if a geographical phenomenon has no characteristic scale, its length, area, volume, eigenvalue, average value, or standard deviation will depend measurement scale or sample size. The probability density distribution of a scale-free phenomenon always takes on a long-tailed curve, which cannot be converted into a unimodal curve.

The geographical phenomena in the real world are complex. For a lake, its boundary line has no characteristic length, but the area within the boundary has characteristic length, which can be represented by the radius of the equivalent circle of the lake's area. Where a city is concerned, urban population density distribution bears characteristic scale and can be described by Clark's model (Clark, 1951), but the urban traffic network density distribution has no characteristic scale and should be described by Smeed's model (Smeed, 1963; Batty and Longley, 1994). In practice, we should adopt appropriate mathematical methods for data processing, mathematical modeling and quantitative analysis according to different properties of geographical phenomena (Table 5). During the quantitative revolution of geography, the development of geographical science once made remarkable achievements. Unfortunately, due to the limited conditions at that time, the geographers could not distinguish scale-free phenomena from scaleful geographical phenomena. Many geographic models could not be used to make proper explanation and prediction. As a result, the theorization of geography suffered setbacks.

**Table 5 Two types of geographical space and the corresponding mathematical methods**

| Space | Model | Parameter | Mathematics tool |
|---|---|---|---|
| **Distance-based space** | Based on exponential decay, logarithmic decay, normal decay, lognormal decay, gamma decay, etc. | Characteristic length (characteristic distance, or characteristic radius) | Euclidean geometry, higher mathematics including calculus, linear algebra, probability theory and statistics |
| **Dimension-based space** | Based on power law decay | Scaling exponent, esp., fractal dimension | Fractal geometry, allometric theory, complex network theory, wavelet analysis, renormalization group, etc. |

## 4.2 Three types of geographical space

For geographical phenomena with characteristic scales, mathematical tools have been well developed. Conventional advanced mathematics and the spatial statistics based on advanced mathematics are enough for geographical to make spatial analyses. However, for the geographical phenomena without characteristic scale, we need new concepts and new methods for mathematical description and geographical explanation. Many mathematical methods such as fractal geometry, complex network theory, wavelet analysis, and renormalization group can be employed to make scaling analyses of geographical systems. Among various mathematical methods, fractal geometry is the most effective one for scale-free research on geographical phenomena. The basic and important parameter of fractal analysis is fractal dimension. Different types of fractal dimensions have different uses in scale geographical spatial analyses. The most critical problem is to distinguish different geographical spaces from one another, so as to choose effective methods and fractal dimensions for spatial description and explanation.

According to the processes of measurements and calculations of fractal dimension, geographical space can be divided into three types. The first one is the real space (R-space for short), the second one is the phase space (P-space for short), and the third one is the order space (O-space for short) (Chen, 2008; Chen, 2014). The real space is the first geographical space, which is easiest to understand. Such space can be surveyed through field, maps, and remote sensing images. The models for the real space are based on spatial data from spatial measurements, census, and sampling. The phase space and order space are relatively abstract and difficult to be understood. The phase space is the second geographical space, which can be described by time series data of geographical evolution. The order space is third geographical space, which can be characterized by cross-sectional data or hierarchical series data (Table 6). Where the classic geographical mathematical models are concerned, the urban density decay models (e.g., Clark's model, Smeed's model) are defined in the real space, the longitudinal allometric growth (e.g., the urban area-population allometry based on dynamic evolution) models are defined in the phase space, and the rank-size rule (e.g. Zipf's law) are defined in order space. Allometric scaling models fall into three categories. The spatial allometry belongs to real space, the longitudinal allometry belongs to phase space, and the cross-sectional allometry belong to order space (Chen, 2014; Chen *et al*, 2019).

**Table 6 Three types of spatial concepts for geographical analyses**

| Type | Object | Data | Model | Fractal dimension |
| --- | --- | --- | --- | --- |
| Real space (R-space) | Spatial patterns | Spatial data | Spatial allometry, equation (8) | Box dimension |
| Phase space (P-space) | Dynamic process | Time series data | Temporal allometry, equation (11) | Correlation dimension |
| Order space (O-space) | Hierarchical structure | Cross-sectional data | Hierarchical allometry, equation (12) | Similarity dimension |

Why should we divide geospatial space into three types? The reason is simple, that is, we have three types of geographical observational data. Spatial data, time series data, and cross-section data can generate different types of fractal dimensions for the correlated geographical phenomena. The fractal parameters based on time series data and cross-section data cannot be attributed to real space. Moreover, the fractal dimension based on time series is not the same as that based on cross-sectional data. These two types of fractal parameters should not belong to the same type of geographical space. Because of the confusion of space types, the explanation of fractal parameters used to confused with each other. For example, the fractal dimension based on Horton-Strahler's law of river composition is actually the fractal parameter of order space, but it is confused with that of real space of river systems. Geomorphologists can't explain the phenomenon that the fractal dimension is sometimes greater than 2 or less than 1 (LaBarbera and Rosso, 1989; Rosso *et al*, 1991). The expected fractal dimension values come between 1 and 2. In fact, the fractal dimension of real space is supposed to vary from 1 to 2. The fractal dimension for real space should be estimated with box-counting method. The law of river composition proposed by Horton (1945) and developed by Strahler (1952) can be used to estimated the similarity dimension of river systems, and the similarity dimension is the fractal dimension defined in order space and is not always consistent with the fractal dimension defined in real space. Similarly, for the central place systems, the fractal dimension based on spatial network structure differs from but links to the fractal dimension based on hierarchical structure (Chen, 2008).

Fractal dimension and the related scaling exponents compose the main parameter sets of spatial dimension for scale-free geographical systems. Using different methods, we can obtain different types of fractal dimensions and scaling exponents. There are various approaches to measuring fractal dimensions values (Batty and Longley, 1994; Chen, 2019; Frankhauser, 1994; Frankhauser, 1998; Takayasu, 1990). Therefore, fractal parameters are diverse, and different fractal dimensions have different uses for spatial analysis (Table 7). The key is to select the appropriate fractal dimension estimation methods for different research objectives and objects. For the geographical phenomena in the real space, we can make use of box-counting method, growing cluster method (radial method), sandbox method, divider method, and so on. For the geographical processes in the phase space, we can calculate the correlation dimension by reconstructing phase space using time series, or estimate the longitudinal allometric scaling exponent by means of a pair time series. For the geographical systems in the order space, we can estimate the similarity dimension by means of rank-size distribution or hierarchical structure. The fractal dimensions of these three spaces are different, but they corresponding relations to one another (Chen, 2014).

**Table 7 Fractal dimension and scaling exponents for geographical analyses in dimension-based space**

| Space | Method | Fractal dimension | Use |
|---|---|---|---|
| Real space (R-space) [Spatial data, digital maps, remoted sensing images] | Box counting method | Box dimension | Spatial distribution |
| | Prism counting method | Prism box dimension | Spatial distribution |
| | Sandbox method | Sandbox dimension | Growth process |
| | Spatial correlation analysis | Spatial correlation dimension | Spatial structure |
| | Area/Number-radius scaling (cluster growing method) | Radial dimension | Growth process |
| | Wave spectrum scaling | Form dimension | Growth process |
| | Walking-divider method | Boundary dimension | Geographical line |
| | Perimeter-area scaling | Boundary dimension | Geographical line |

|  | …… | …… | …… |
|---|---|---|---|
| Phase space (P-space) [Time series data] | Reconstructing phase space | Correlation dimension | Dynamic process |
| | Elasticity relation | Similarity dimension | Dynamic relation |
| | Power spectrum scaling | Self-affine record dimension | Growth |
| | …… | …… | …… |
| Order space (O-space) [Cross-sectional data, rank-size series, hierarchical series] | Pareto distribution | Similarity dimension | Size distribution |
| | Hierarchical scaling | Similarity dimension | Hierarchical structure |
| | Renormalization | Similarity dimension | Network structure |
| | Allometric scaling | Similarity dimension ratio | Relative growth |
| | …… | …… | …… |

# 5 Conclusions

Geographers have developed many good mathematical models, analytical methods and spatial statistical technique for the geographical phenomena with characteristic scales. However, for the geographical phenomena without characteristic scale, the methodology of spatial analyses were less-developed and have been developing these years. The aim of this paper is to clarify the differences and connections between the two kinds of geographical spatial analyses. The mains viewpoints of this paper can be summarized as follows. **First, the space concept for geographical analysis should be divided into two types: distance-based space and dimension-based space.** This classification is helpful for geographers to choose proper analytical methods for specific problems. Geographical phenomena fall into two categories: scaleful phenomena and scale-free phenomena. The former can be modeled and analyzed in distance-based space and the latter should be modeled and analyzed in dimension-based space. The geographical spatial analysis for scaleful phenomena is based on typical distance. However, if a geographical phenomenon has no characteristic scale, simple distance variable no longer guarantees the validity of spatial analysis. In this case, we should make scaling analysis based on variable distance, and thus the distance-based space is actually replaced by dimension-based space. **Second, the scaleful geographical phenomena are defined in distance-based space and can be modeled and analyzed using** **conventional mathematical methods.** If a geographical phenomenon bear characteristic scales, it can be modeled and quantitatively analyzed by means of higher mathematics method and spatial statistics method based on higher mathematics. In this case, a typical distance can be used as an effective spatial characteristic value. Based on distance or distance matrix, varied mathematical models of geographical systems can be built. Data processing and quantitative analysis can be made on the base of these models. This type of geographical mathematical methods has been well developed so far. **Third, the scale-free geographical phenomena are defined in dimension-based space and should be modeled and analyzed by the mathematical tools based on scaling idea.** If a geographical phenomenon has no characteristic scale, we will be unable to find characteristic distance, and conventional geographical mathematical methods will be invalid. Using the power law relationships between distance and corresponding spatial measures, we can calculate fractal dimension or scaling exponent. A scaling exponent is a ratio of two fractal dimensions of a function of a fractal dimension. Fractal dimension and scaling exponents are basic parameters for dimension-based spatial analysis. According to different nature of fractal dimension, dimension-based geographical space can be divided into three types: real space (based on geographical landscape), phase space (based on time series), and order space (based on hierarchical structure). The three spaces correspond to spatial data, time series data, and cross-sectional data, respectively.

## Acknowledgements

This research was sponsored by the National Natural Science Foundations of China (Grant No. 41671167). The support is gratefully acknowledged.